# Anisotropic Multi-layer Elliptical Waveguides Incorporating Graphene Layers: A Novel Analytical Model

Mohammad Bagher Heydari[1,*], Mohammad Hashem Vadjed Samiei[1]

[1,*]School of Electrical Engineering, Iran University of Science and Technology (IUST), Tehran, Iran

**This article aims to propose a novel analytical model for anisotropic multi-layer elliptical structures incorporating graphene layers. The multi-layer structure is formed of various magnetic materials, each one has the permittivity and permeability tensors of $\bar{\bar{\varepsilon}}, \bar{\bar{\mu}}$, respectively. An external magnetic bias has been applied in the axial direction. A graphene layer, with isotropic surface conductivity of $\sigma$, has been sandwiched between two adjacent anisotropic materials. A novel matrix representation has been derived to find the propagation parameters of the multi-layer structure. Two exemplary important cases of the proposed general structure, as waveguides, have been investigated to show, first the validity of our proposed analytical model, and second, the richness of the general structure. The analytical and simulation results show an excellent agreement. A very large value of figure of merit (FOM), e.g. FOM=110, is achieved for the second structure for the chemical potential and external magnetic bias of $\mu_c = 0.9 \, eV$ and $B_0 = 1 \, T$, respectively. Our general structure and its analytical model can be exploited to design innovative THz devices such as absorbers, couplers, and cloaks.**

## I. INTRODUCTION

GRAPHENE has opened a new emerging science between scientists, which is known as "Graphene Plasmonics" in the literature [1]. This novel science is based on the usage of graphene plasmon polaritons (GPPs) to design innovative and tunable THz devices such as such as waveguides [2-13], isolator [14], circulator [15, 16], coupler [17], resonator [18], antennas [19-21], filter [22], Radar Cross-Section (RCS) reduction-based devices [23-25], and graphene-based medical components [26-32]. It should be noted that noble metals support SPPs at the near-infrared and visible frequencies [18, 33-42].

Among the various types of graphene-based devices, elliptical structures have attracted the attention of scientists due to their fascinating applications, such as absorbers [43, 44], waveguides [45, 46] and cloaks [47]. In [45], the modal properties of a graphene-coated elliptical nano-wire has been studied by utilizing the finite element method (FEM). A normalized mode area of $10^{-4} \sim 10^{-3}$ and long propagation length (about 285μm) has been reported in [45]. X. Cheng et.al have investigated the modal properties of elliptical nano-wire with double-layer graphene in [46]. They have reported a value of 3.24μm for the propagation length of the fundamental mode in this waveguide [46].

To the authors' knowledge, a comprehensive study on anisotropic multi-layer elliptical waveguides with graphene layers has not been reported and studied in any published article. To cover all special cases of graphene-based elliptical waveguides, we propose a new theoretical model, for a general multilayer elliptical structure, by using the "separation of variables" method. Our general structure is composed of various magnetic materials, where a graphene layer has been sandwiched between two adjacent anisotropic layers. Each magnetic material has the permittivity and permeability tensors of $\bar{\bar{\varepsilon}}, \bar{\bar{\mu}}$, respectively. An external magnetic field has been applied in the z-direction. Graphene layers have isotropic surface conductivity, known as "Kubo's formula" in the literature [48]. Our complete analysis obtains the closed-form relations of field distributions and propagating parameters for all plasmonic modes. The numerical methods in electromagnetics are volumetric and they solve a problem by discretizing the structure. Hence, there is no computational method faster than the analytical model based on closed-form relations. In summary, our analytical model has two major advantages: first, it has high accuracy, and second, it can compute the plasmonic parameters very fast, compared to computational methods. This rich multi-layer structure and its accurate analytical model helps one to design tunable THz devices. The propagating plasmonic features of these devices are adjustable by altering the external magnetic bias and the chemical potential of the graphene. Furthermore, having two geometric design parameters in elliptical structures, i.e. the long axis and short axis of each ellipse, more degrees of freedom are available to the designer. Hence, compared to graphene-based cylindrical waveguides, a superior performance (the high value of FOM) is achievable in these structures.

This paper is organized as follows. Section II presents the proposed general structure and its novel analytical model. The model will be derived by using Mathieu functions [49] in elliptical coordinates inside bi-gyrotropic media. In this section, a matrix representation will be obtained for the general structure. Two exemplary structures are investigated in section III. The first, familiar structure, is a graphene-based elliptical

*Corresponding author: Mohammad Bagher Heydari (e-mail: mo_heydari@alumni.iust.ac.ir )

nano-wire, which will be studied to show the validity of our proposed model. Then, a novel anisotropic structure is introduced and investigated to indicate the richness of the proposed general structure regarding the related specific plasmonic wave phenomena and effects. To the authors' knowledge, applying the gyro-electric layer in the graphene-based elliptical waveguide has not been reported in any research article. The hybridization of graphene and the gyro-electric layer produces hybrid plasmonic waves, which their plasmonic parameters are adjustable by changing the magnetic bias and the chemical potential. Furthermore, a very large value of FOM, e.g. FOM=110 for $\mu_c = 0.9\ eV, B_0 = 1\ T$, is achieved for the second structure. Finally, section IV concludes the article.

## II. THE PROPOSED GENERAL STRUCTURE AND ITS ANALYTICAL MODEL

Fig. 1 illustrates the schematic of the proposed structure, where each graphene layer has been sandwiched between two adjacent anisotropic materials, each material has the permittivity and permeability tensors of $\bar{\bar{\varepsilon}}$ and $\bar{\bar{\mu}}$, respectively. To excite the plasmonic modes, the electric and magnetic currents have been considered outside the structure, seen as "red dashed lines" in Fig. 1. It should be noted that one of these currents is sufficient for exciting plasmonic waves. Nevertheless, we have considered both of them to generalize our analysis. An external magnetic bias ($B_0$) has been applied in the z-direction. The conductivity of the graphene layer is modeled by Kubo's formula since the DC magnetic bias is parallel to the graphene [48]:

$$\sigma_N(\omega, \mu_{c,N}, \Gamma_N, T) = \frac{-je^2}{4\pi\hbar} Ln\left[\frac{2|\mu_{c,N}| - (\omega - j2\Gamma_N)\hbar}{2|\mu_{c,N}| + (\omega - j2\Gamma_N)\hbar}\right] + \frac{-je^2 K_B T}{\pi\hbar^2(\omega - j2\Gamma_N)}\left[\frac{\mu_{c,N}}{K_B T} + 2Ln\left(1 + e^{-\mu_{c,N}/K_B T}\right)\right] \quad (1)$$

Where $h$ is the reduced Planck's constant, $K_B$ is Boltzmann's constant, $\omega$ is radian frequency, $e$ is the electron charge, $\Gamma_N$ is the phenomenological electron scattering rate for the N-th layer ($\Gamma_N = 1/\tau_N$, where $\tau_N$ is the relaxation time), $T$ is the temperature, and $\mu_{c,N}$ is the chemical potential for the N-th layer which can be altered by chemical doping or electrostatic bias [48].

In general, the permittivity and permeability tensors of the N-th layer for a DC magnetic bias in the z-direction are written as following tensors [50]:

$$\bar{\bar{\mu}}_N = \mu_0 \begin{pmatrix} \mu_N & j\mu_{a,N} & 0 \\ -j\mu_{a,N} & \mu_N & 0 \\ 0 & 0 & \mu_{\parallel,N} \end{pmatrix} \quad (2)$$

$$\bar{\bar{\varepsilon}}_N = \varepsilon_0 \begin{pmatrix} \varepsilon_N & j\varepsilon_{a,N} & 0 \\ -j\varepsilon_{a,N} & \varepsilon_N & 0 \\ 0 & 0 & \varepsilon_{\parallel,N} \end{pmatrix} \quad (3)$$

Where $\varepsilon_0$ and $\mu_0$ are the permittivity and permeability of the free space, respectively.

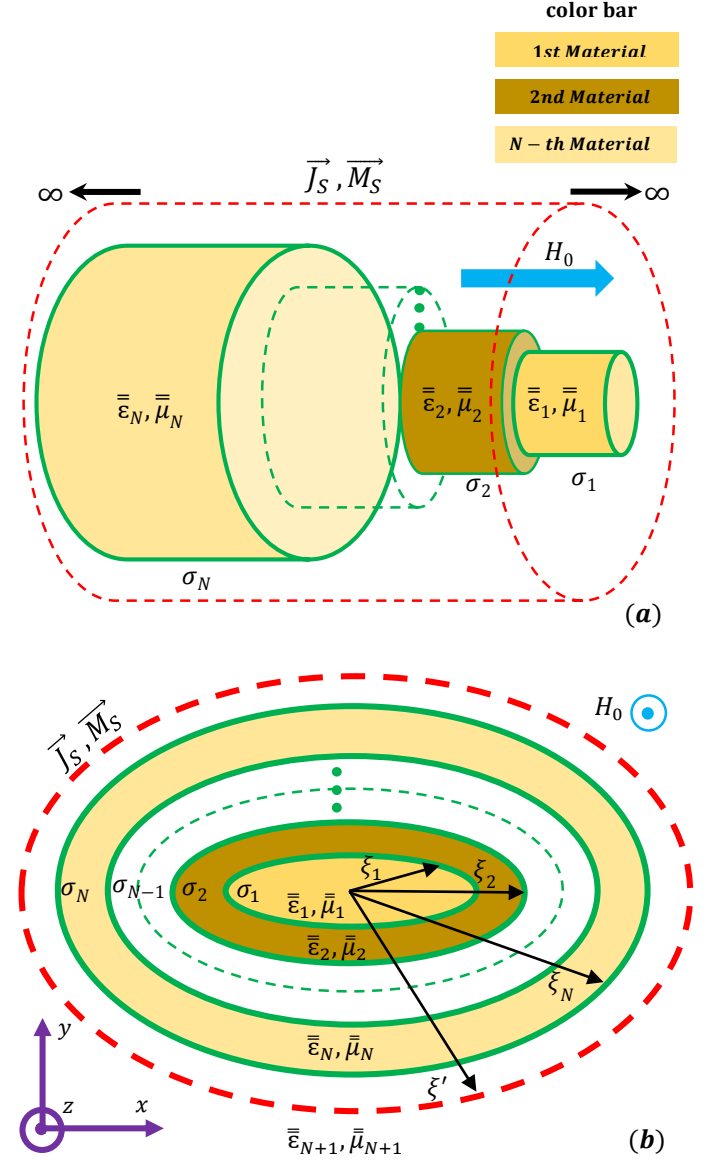

Fig. 1. (a) The 3D schematic of a general anisotropic multi-layer elliptical structure containing electrically biased graphene layers, (b) The cross-section schematic of the general structure. The structure is infinite in the z-direction.

In (2), the diagonal and off-diagonal elements are expressed as [50]:

$$\mu_N = 1 + \frac{\omega_{M,N}(\omega_{H,N} + j\omega\alpha_N)}{\omega_{H,N}^2 - (1 + \alpha_N^2)\omega^2 + 2j\alpha_N\omega\omega_{H,N}} \quad (4)$$

$$\mu_{a,N} = \frac{\omega \omega_{M,N}}{\omega_{H,N}^2 - (1+\alpha_N^2)\omega^2 + 2j\alpha_N \omega \omega_{H,N}} \quad (5)$$

$$\mu_{\|,N} = 1 - \frac{j\alpha_N \omega_{M,N}}{\omega + j\alpha_N \omega_{H,N}} \quad (6)$$

Where $\gamma_N$ and $\alpha_N$ are the gyromagnetic ratio and the Gilbert damping constant for the $N$-th layer, respectively. Moreover, $\omega_{H,N} = \gamma_N H_0$, $\omega_M = \gamma_N M_S$ and $M_S$ is the saturation magnetization.

The diagonal and off-diagonal elements of permittivity tensors have well-known forms [51]:

$$\varepsilon_N = \varepsilon_{\infty,N} \left(1 - \frac{\omega_{p,N}^2 (\omega + j\upsilon_N)}{\omega\left[(\omega + j\upsilon_N)^2 - \omega_{c,N}^2\right]}\right) \quad (7)$$

$$\varepsilon_{a,N} = \varepsilon_{\infty,N} \left(\frac{\omega_{p,N}^2 \omega_c}{\omega\left[(\omega + j\upsilon_N)^2 - \omega_c^2\right]}\right) \quad (8)$$

$$\varepsilon_{\|,N} = \varepsilon_{\infty,N} \left(1 - \frac{\omega_{p,N}^2}{\omega(\omega + j\upsilon_N)}\right) \quad (9)$$

Where $\nu_N$ and $\varepsilon_{\infty,N}$ are the effective collision rate and the background permittivity of the $N$-th layer, respectively. Furthermore,

$$\omega_{p,N} = \sqrt{\frac{n_s e^2}{\varepsilon_0 \varepsilon_{\infty,N} m^*}} \quad (10)$$

$$\omega_c = \frac{e B_0}{m^*} \quad (11)$$

are the plasma and cyclotron frequencies, respectively. In these relations, $e, m^*$ and $n_s$ are the charge, effective mass, and the density of the carriers.

To analyze the general structure, Mathieu functions must be used [49]. The fields should be written in the elliptical coordinates $(\xi, \eta)$ and the boundary conditions are applied to obtain the dispersion relation. It should be noted that the propagation is supposed in the z-direction. Before embarking the theoretical analysis of the proposed structure, let us briefly review the general mathematical relations in the elliptical coordinates. The following relations are defined in the elliptical cylinder coordinates in the terms of the rectangular coordinates $(x, y, z)$ [49]:

$$\begin{aligned} x &= f \cosh\xi \cdot \cos\eta \\ y &= f \sinh\xi \cdot \sin\eta \\ z &= z \quad (0 \le \xi < \infty, \; 0 < \eta < 2\pi) \end{aligned} \quad (12)$$

Where $f$ is the semi-focal length of the ellipse. Consider Maxwell's equations inside the $N$-th layer in the frequency domain (suppose $e^{i\omega t}$) [50]:

$$\nabla \times \mathbf{E} = -j\omega \bar{\bar{\mu}}_N \cdot \mathbf{H} \quad (13)$$

$$\nabla \times \mathbf{H} = j\omega \bar{\bar{\varepsilon}}_N \cdot \mathbf{E} \quad (14)$$

By doing some mathematical procedures, the z-components of the electric and magnetic field inside the bi-gyrotropic layer are expressed [50]:

$$\left(\nabla_\perp^2 + \frac{\varepsilon_{\|,N}}{\varepsilon_N}\frac{\partial^2}{\partial z^2} + (k_0^2 \varepsilon_{\|,N} \mu_{\perp,N})\right) E_{z,N} + \\ k_0 \mu_{\|,N}\left(\frac{\varepsilon_{a,N}}{\varepsilon_N} + \frac{\mu_{a,N}}{\mu_N}\right)\frac{\partial}{\partial z} H_{z,N} = 0 \quad (15)$$

$$\left(\nabla_\perp^2 + \frac{\mu_{\|,N}}{\mu_N}\frac{\partial^2}{\partial z^2} + (k_0^2 \varepsilon_{\perp,N} \mu_{\|,N})\right) H_{z,N} - \\ k_0 \varepsilon_{\|,N}\left(\frac{\varepsilon_{a,N}}{\varepsilon_N} + \frac{\mu_{a,N}}{\mu_N}\right)\frac{\partial}{\partial z} E_{z,N} = 0 \quad (16)$$

where

$$\nabla_\perp^2 = \frac{1}{h}\left(\frac{\partial^2}{\partial^2 \xi} + \frac{\partial^2}{\partial^2 \eta}\right) \quad (17)$$

In these equations, $k_0$ is the free space wave-number and,

$$\varepsilon_{\perp,N} = \varepsilon_N - \frac{\varepsilon_{a,N}^2}{\varepsilon_N} \quad (18)$$

$$\mu_{\perp,N} = \mu_N - \frac{\mu_{a,N}^2}{\mu_N} \quad (19)$$

$$h = f\sqrt{\cosh^2\xi - \cos^2\eta} \quad (20)$$

In general form in the elliptical coordinates, the z-component of the electric and magnetic fields for the plasmonic waves propagating in the z-direction can be written as:

$$H_{z,N}(\xi,\eta,z) = \int_{-\infty}^{+\infty} \sum_{m=0}^{\infty} e^{jk_z z} H_{z,N}^m(\xi,\eta;q)\, dk_z \quad (21)$$

$$E_{z,N}(\xi,\eta,z) = \int_{-\infty}^{+\infty} \sum_{m=0}^{\infty} e^{jk_z z} E_{z,N}^m(\xi,\eta;q)\, dk_z \quad (22)$$

Where $m$ is the order of Mathieu functions and $H_{z,N}^m, E_{z,N}^m$ are the potential functions in the elliptical coordinates [49]:

$$\begin{cases} H_{z,N}^m(\xi,\eta;q) \\ E_{z,N}^m(\xi,\eta;q) \end{cases} = R_m(\xi;q) \cdot \Phi_m(\eta;q) \quad (23)$$

$R_m(\xi, q)$ is oscillatory radial Mathieu function of the first and second kind (for $q > 0$) [49]:

$$R_m(\xi;q) = \begin{cases} Je_m(\xi,q),\ Jo_m(\xi,q) & \text{first kind} \\ Ne_m(\xi,q),\ No_m(\xi,q) & \text{second kind} \end{cases} \quad (24)$$

and for $q < 0$, evanescent Mathieu functions are appeared [49]:

$$R_m(\xi;q) = \begin{cases} Ie_m(\xi,q),\ Io_m(\xi,q) & \text{first kind} \\ Ke_m(\xi,q),\ Ko_m(\xi,q) & \text{second kind} \end{cases} \quad (25)$$

Radial Mathieu functions can be odd or even, where "e" and "o" indicate even and odd functions in the above relations, respectively. Also, even and odd angular Mathieu functions are expressed as [49]:

$$\Phi_m(\eta;q) = \begin{cases} ce_m(\eta,q) & m = 0,1,2,\ \ldots \\ se_m(\eta,q) & m = 1,2,3,\ \ldots \end{cases} \quad (26)$$

Now, by substituting (21) and (22) into (15) and (16), one can obtain

$$\left(\nabla_\perp^2 + k_0^2 \varepsilon_{\parallel,N} \mu_{\perp,N} - \frac{\varepsilon_{\parallel,N}}{\varepsilon_N} k_z^2\right) E_{z,N}^m - jk_0 k_z \mu_{\parallel,N} \left(\frac{\varepsilon_{a,N}}{\varepsilon_N} + \frac{\mu_{\alpha,N}}{\mu_N}\right) H_{z,N}^m = 0 \quad (27)$$

$$\left(\nabla_\perp^2 + k_0^2 \varepsilon_{\perp,N} \mu_{\parallel,N} - \frac{\mu_{\parallel,N}}{\mu_N} k_z^2\right) H_{z,N}^m + jk_0 k_z \varepsilon_{\parallel,N} \left(\frac{\varepsilon_{a,N}}{\varepsilon_N} + \frac{\mu_{\alpha,N}}{\mu_N}\right) E_{z,N}^m = 0 \quad (28)$$

Combining (27) and (28) results in a fourth-order differential equation:

$$\left(\nabla_\perp^2 + S_{2N-1}^2\right)\left(\nabla_\perp^2 + S_{2N}^2\right) H_{z,N}^m = 0 \quad (29)$$

Then, the characteristics equation of (29) for the N-th layer of the anisotropic medium is obtained

$$s^4 + G_{1,N} s^2 + G_{2,N} = 0 \quad (30)$$

with the following coefficients,

$$G_{1,N} = -k_0^2\left(\varepsilon_{\parallel,N}\mu_{\perp,N} + \varepsilon_{\perp,N}\mu_{\parallel,N}\right) + \left(\frac{\varepsilon_{\parallel,N}}{\varepsilon_N} + \frac{\mu_{\parallel,N}}{\mu_N}\right) k_z^2 \quad (31)$$

$$G_{2,N} = \left(k_0^2 \varepsilon_{\parallel,N}\mu_{\perp,N} - \frac{\varepsilon_{\parallel,N}}{\varepsilon_N} k_z^2\right)\left(k_0^2 \varepsilon_{\perp,N}\mu_{\parallel,N} - \frac{\mu_{\parallel,N}}{\mu_N} k_z^2\right) - k_0^2 k_z^2 \mu_{\parallel,N}\varepsilon_{\parallel,N}\left(\frac{\varepsilon_{a,N}}{\varepsilon_N} + \frac{\mu_{\alpha,N}}{\mu_N}\right)^2 \quad (32)$$

If we suppose that the roots of (30) for the N-th layer are $S_{2N-1}, S_{2N}$, then the following parameters will be defined:

$$q_{2N-1} = \frac{f^2}{4} S_{2N-1} \quad (33)$$

$$q_{2N} = \frac{f^2}{4} S_{2N} \quad (34)$$

Hence, the roots of characteristics equations for various regions of Fig.1 are assumed

$$q = \begin{cases} q_1, q_2 & i = 1, 2\ ;\ \text{for first layer} \\ \ldots & \ldots \\ q_{2N-1}, q_{2N} & i = 2N-1, 2N\ ;\ \text{for } N\text{-th layer} \\ q_{2N+1}, q_{2N+2} & i = 2N+1, 2N+2\ ;\ \text{for } N+1\text{-th layer} \end{cases} \quad (35)$$

In the above relations, $N$ indicates the number of the layer and $i$ shows the index of the roots for that layer. Now, electromagnetic waves in various regions should be written. For $\xi < \xi_1$, one can write:

$$H_{z,1}^m(\xi,\eta;q) = A_{m,1,1} Je_m(\xi,q_1) ce_m(\eta,q_1) + A'_{m,1,1} Jo_m(\xi,q_1) se_m(\eta,q_1) + A_{m,2,1} Je_m(\xi,q_2) ce_m(\eta,q_2) + A'_{m,2,1} Jo_m(\xi,q_2) se_m(\eta,q_2) \quad (36)$$

$$E_{z,1}^m(\xi,\eta;q) = T_{1,1}\begin{pmatrix} A'_{m,1,1} Je_m(\xi,q_1) ce_m(\eta,q_1) + \\ A_{m,1,1} Jo_m(\xi,q_1) se_m(\eta,q_1) \end{pmatrix} + T_{2,1}\begin{pmatrix} A'_{m,2,1} Je_m(\xi,q_2) ce_m(\eta,q_2) + \\ A_{m,2,1} Jo_m(\xi,q_2) se_m(\eta,q_2) \end{pmatrix} \quad (37)$$

For $\xi_1 < \xi < \xi_2$:

$$H_{z,2}^m(\xi,\eta;q) = ce_m(\eta,q_3)\left[A_{m,3,2} Je_m(\xi,q_3) + B_{m,3,2} Ne_m(\xi,q_3)\right] + se_m(\eta,q_3)\left[A'_{m,3,2} Jo_m(\xi,q_3) + B'_{m,3,2} No_m(\xi,q_3)\right] + ce_m(\eta,q_4)\left[A_{m,4,2} Je_m(\xi,q_4) + B_{m,4,2} Ne_m(\xi,q_4)\right] + se_m(\eta,q_4)\left[A'_{m,4,2} Jo_m(\xi,q_4) + B'_{m,4,2} No_m(\xi,q_4)\right] \quad (38)$$

$$E_{z,2}^m(\xi,\eta;q) = T_{3,2}\, ce_m(\eta,q_3)\left[A'_{m,3,2} Je_m(\xi,q_3) + B'_{m,3,2} Ne_m(\xi,q_3)\right] + T_{3,2}\, se_m(\eta,q_3)\left[A_{m,3,2} Jo_m(\xi,q_3) + B_{m,3,2} No_m(\xi,q_3)\right] + T_{4,2}\, ce_m(\eta,q_4)\left[A'_{m,4,2} Je_m(\xi,q_4) + B'_{m,4,2} Ne_m(\xi,q_4)\right] + T_{4,2}\, se_m(\eta,q_4)\left[A_{m,4,2} Jo_m(\xi,q_4) + B_{m,4,2} No_m(\xi,q_4)\right] \quad (39)$$

In general, for $N$-th layer ($\xi_{N-1} < \xi < \xi_N$):

$$H_{z,N}^m(\xi,\eta;q) =$$
$$ce_m(\eta,q_{2N-1})\left[A_{m,2N-1,N}Je_m(\xi,q_{2N-1}) + B_{m,2N-1,N}Ne_m(\xi,q_{2N-1})\right] +$$
$$se_m(\eta,q_{2N-1})\left[A'_{m,2N-1,N}Jo_m(\xi,q_{2N-1}) + B'_{m,2N-1,N}No_m(\xi,q_{2N-1})\right] +$$
$$ce_m(\eta,q_{2N})\left[A_{m,2N,N}Je_m(\xi,q_{2N}) + B_{m,2N,N}Ne_m(\xi,q_{2N})\right] +$$
$$se_m(\eta,q_{2N})\left[A'_{m,2N,N}Jo_m(\xi,q_{2N}) + B'_{m,2N,N}No_m(\xi,q_{2N})\right]$$
(40)

$$E_{z,N}^m(\xi,\eta;q) =$$
$$T_{2N-1,N}\, ce_m(\eta,q_{2N-1})\begin{bmatrix} A'_{m,2N-1,N}Je_m(\xi,q_{2N-1}) + \\ B'_{m,2N-1,N}Ne_m(\xi,q_{2N-1}) \end{bmatrix} +$$
$$T_{2N-1,N}\, se_m(\eta,q_{2N-1})\begin{bmatrix} A_{m,2N-1,N}Jo_m(\xi,q_{2N-1}) + \\ B_{m,2N-1,N}No_m(\xi,q_{2N-1}) \end{bmatrix} +$$ (41)
$$T_{2N,N}\, ce_m(\eta,q_{2N})\begin{bmatrix} A'_{m,2N,N}Je_m(\xi,q_{2N}) + \\ B'_{m,2N,N}Ne_m(\xi,q_{2N}) \end{bmatrix} +$$
$$T_{2N,N}\, se_m(\eta,q_{2N})\begin{bmatrix} A_{m,2N,N}Jo_m(\xi,q_{2N}) + \\ B_{m,2N,N}No_m(\xi,q_{2N}) \end{bmatrix}$$

Finally, for $\xi > \xi'$:
$$H_{z,\xi>\xi'}^m(\xi,\eta;q) =$$
$$C_{m,2N+1,N+1}Ne_m(\xi,q_{2N+1})ce_m(\eta,q_{2N+1}) +$$
$$C'_{m,2N+1,N+1}No_m(\xi,q_{2N+1})se_m(\eta,q_{2N+1}) +$$ (42)
$$D_{m,2N+2,N+1}Ne_m(\xi,q_{2N+2})ce_m(\eta,q_{2N+2}) +$$
$$D'_{m,2N+2,N+1}No_m(\xi,q_{2N+2})se_m(\eta,q_{2N+2})$$

$$E_{z,\xi>\xi'}^m(\xi,\eta;q) =$$
$$T_{2N+1,N+1}\begin{bmatrix} C'_{m,2N+1,N+1}Ne_m(\xi,q_{2N+1})ce_m(\eta,q_{2N+1}) + \\ C_{m,2N+1,N+1}No_m(\xi,q_{2N+1})se_m(\eta,q_{2N+1}) \end{bmatrix} +$$
$$T_{2N+2,N+1}\begin{bmatrix} D'_{m,2N+2,N+1}Ne_m(\xi,q_{2N+2})ce_m(\eta,q_{2N+2}) + \\ D_{m,2N+2,N+1}No_m(\xi,q_{2N+2})se_m(\eta,q_{2N+2}) \end{bmatrix}$$
(43)

In (36)-(43), $T_{i,N}$ are defined as:
$$T_{i,N} = \frac{q_i^2 - k_0^2 \varepsilon_{\perp,N}\mu_{\parallel,N} + \dfrac{\mu_{\parallel,N}}{\mu_N}k_z^2}{-jk_0 k_z \varepsilon_{\parallel,N}\left(\dfrac{\varepsilon_{a,N}}{\varepsilon_N} + \dfrac{\mu_{a,N}}{\mu_N}\right)}$$ (44)

$i = 2N-1, 2N$ ; $N = 1,2,...$

In the above relations, $A_{m,1,1}, \acute{A}_{m,1,1}, B_{m,3,2}, \acute{B}_{m,3,2}, ...$ are unknown coefficients and will be determined by applying boundary conditions. It should be noted that the $m$-indices in (36)-(43) show the order of plasmonic modes. For instance, $H_{z,N}(\xi,\eta,z)$ for $N$-th layer $(\xi_{N-1} < \xi < \xi_N)$ is written as follows:

$$H_{z,N}(\xi,\eta,z) =$$
$$\int_{-\infty}^{+\infty}\left[\sum_{m=0}^{\infty}\left(ce_m(\eta,q_{2N-1})\begin{bmatrix} A_{m,2N-1,N}Je_m(\xi,q_{2N-1}) + \\ B_{m,2N-1,N}Ne_m(\xi,q_{2N-1}) \end{bmatrix} + ce_m(\eta,q_{2N})\begin{bmatrix} A_{m,2N,N}Je_m(\xi,q_{2N}) + \\ B_{m,2N,N}Ne_m(\xi,q_{2N}) \end{bmatrix}\right) + \sum_{m=1}^{\infty}\left(se_m(\eta,q_{2N-1})\begin{bmatrix} A'_{m,2N-1,N}Jo_m(\xi,q_{2N-1}) + \\ B'_{m,2N-1,N}No_m(\xi,q_{2N-1}) \end{bmatrix} + se_m(\eta,q_{2N})\begin{bmatrix} A'_{m,2N,N}Jo_m(\xi,q_{2N}) + \\ B'_{m,2N,N}No_m(\xi,q_{2N}) \end{bmatrix}\right)\right]$$
$$\cdot e^{jk_z z}dk_z$$
(45)

The transverse components of electric and magnetic fields are expressed as:

$$\begin{pmatrix} E_{\xi,N} \\ H_{\xi,N} \end{pmatrix} = \frac{1}{h}\left[\bar{\bar{Q}}_N^{Pos}\frac{\partial}{\partial \xi}\begin{pmatrix} E_{z,N} \\ H_{z,N} \end{pmatrix} + j\bar{\bar{Q}}_N^{Neg}\frac{\partial}{\partial \eta}\begin{pmatrix} E_{z,N} \\ H_{z,N} \end{pmatrix}\right]$$ (46)

$$\begin{pmatrix} E_{\varphi,N} \\ H_{\varphi,N} \end{pmatrix} = \frac{1}{h}\left[-j\bar{\bar{Q}}_N^{Neg}\frac{\partial}{\partial \xi}\begin{pmatrix} E_{z,N} \\ H_{z,N} \end{pmatrix} + \bar{\bar{Q}}_N^{Pos}\frac{\partial}{\partial \eta}\begin{pmatrix} E_{z,N} \\ H_{z,N} \end{pmatrix}\right]$$ (47)

Where the Q-matrices in (46) and (47) have been defined:

$$\bar{\bar{Q}}_N^{Pos} = \frac{1}{2}\left[\frac{1}{-k_z^2 + k_0^2\varepsilon_{+,N}\mu_{+,N}}\begin{pmatrix} jk_z & -\omega\mu_0\mu_{+,N} \\ \omega\varepsilon_0\varepsilon_{+,N} & jk_z \end{pmatrix} + \frac{1}{-k_z^2 + k_0^2\varepsilon_{-,N}\mu_{-,N}}\begin{pmatrix} jk_z & \omega\mu_0\mu_{-,N} \\ -\omega\varepsilon_0\varepsilon_{-,N} & jk_z \end{pmatrix}\right]$$
(48)

$$\bar{\bar{Q}}_N^{Neg} = \frac{1}{2}\left[\frac{1}{-k_z^2 + k_0^2\varepsilon_{+,N}\mu_{+,N}}\begin{pmatrix} jk_z & -\omega\mu_0\mu_{+,N} \\ \omega\varepsilon_0\varepsilon_{+,N} & jk_z \end{pmatrix} - \frac{1}{-k_z^2 + k_0^2\varepsilon_{-,N}\mu_{-,N}}\begin{pmatrix} jk_z & \omega\mu_0\mu_{-,N} \\ -\omega\varepsilon_0\varepsilon_{-,N} & jk_z \end{pmatrix}\right]$$
(49)

Moreover,
$$\varepsilon_{\pm,N} = \varepsilon_N \pm \varepsilon_{a,N}$$ (50)

$$\mu_{\pm,N} = \mu_N \pm \mu_{a,N} \quad (51)$$

Now, we should apply boundary conditions to complete the analytical model. In general form, the boundary conditions for the graphene layer sandwiched between two magnetic materials are written as follows:

$$E_{z,N} = E_{z,N+1} \;,\; E_{\eta,N} = E_{\eta,N+1} \quad N = 1,2,3,... \quad (52)$$

$$H_{z,N+1} - H_{z,N} = -\sigma E_{\eta,N} \;,$$
$$H_{\eta,N+1} - H_{\eta,N} = \sigma E_{z,N} \quad N = 1,2,3,... \quad (53)$$

And for the last boundary at $\xi = \hat{\xi}$,

$$E_{z,N+1}^{>} - E_{z,N+1}^{<} = M_{s\eta} \;,\; E_{\eta,N+1}^{>} - E_{\eta,N+1}^{<} = -M_{sz} \quad (54)$$

$$H_{z,N+1}^{>} - H_{z,N+1}^{<} = -J_{s\eta} \;,\; H_{\eta,N+1}^{>} - H_{\eta,N+1}^{<} = J_{sz} \quad (55)$$

In (54) and (55), $M_{sz}, M_{s\eta}, J_{sz}, J_{s\eta}$ are $z$ and $\eta$-components of magnetic and electric currents at $\xi = \hat{\xi}$, respectively. By applying the boundary conditions expressed in (52)-(55), the final matrix representation for our general structure is obtained:

$$\bar{\bar{S}}_{8N+8,8N+8} \cdot \begin{pmatrix} A_{m,1,1} \\ A'_{m,1,1} \\ A_{m,2,1} \\ A'_{m,2,1} \\ A_{m,3,2} \\ A'_{m,3,2} \\ B_{m,3,2} \\ B'_{m,3,2} \\ \vdots \\ C_{m,2N+1,N+1} \\ C'_{m,2N+1,N+1} \\ D_{m,2N+2,N+1} \\ D'_{m,2N+2,N+1} \end{pmatrix}_{8N+8,1} = \begin{pmatrix} 0 \\ 0 \\ 0 \\ 0 \\ 0 \\ 0 \\ 0 \\ 0 \\ \vdots \\ M_{s\eta} \\ -M_{sz} \\ -J_{s\eta} \\ J_{sz} \end{pmatrix}_{8N+8,1} \quad (56)$$

Now, our analytical model has been completed for the general structure of Fig. 1. The matrix $\bar{\bar{S}}$ is a complicated matrix that its elements have been given in the Appendix. This matrix is very important since it obtains the dispersion relation or the propagation constant by setting $det(\bar{\bar{S}}) = 0$. In the next step, the plasmonic parameters of the general multi-layer structure such as the effective index ($n_{eff} = Re(k_z/k_0)$) is straightforward. It should be mentioned that our matrix representation ($\bar{\bar{S}}$) contains both odd ($H_z$ odd symmetry, $E_z$ even symmetry) and even ($H_z$ even symmetry, $E_z$ odd symmetry) modes. Indeed, if we focus on only one type of these modes, the size of the matrix will be decreased to $4N+4$. In what follows, we will consider two exemplary structural variants to show the richness of the proposed general structure.

## III. SPECIAL CASES OF THE PROPOSED STRUCTURE: RESULTS AND DISCUSSIONS

This section studies two special cases of graphene-based elliptical waveguides. These structures are chosen to show, first the validity of the proposed analytical model, and second, the richness of the general structure. The first, familiar, example is a graphene-coated elliptical nano-wire, where a graphene layer has been located on the SiO$_2$ surface. The second waveguide is a novel structure, constituting air-graphene-InSb-SiO$_2$-Si layers. In this waveguide, the gyroelectric layer is the n-type InSb with anisotropic permittivity tensor, which produces tunable GPPs. The propagation parameters of these plasmons can be adjusted via the magnetic bias and the chemical potential. Our study is based just on the first two propagating modes. There are good reasons for relying on just the first two modes. The most important of them: first, the main energy of the propagating wave is carrying by them, and second, the complexity of formulation and hence simulation time will be reduced significantly.

In this paper, the modal analysis (Eigen-value and Eigen-frequency analysis) has been used in our simulations. In our analytical results, we have utilized the determinant of the matrix $\bar{\bar{S}}$ to obtain the propagation parameters. In all simulations, the temperature is supposed to be $T = 300\,K$, the graphene thickness is $\Delta = 0.33\,nm$, and the phenomenological electron scattering rate is $\Gamma = 2 \times 10^{12}\,rad/s$.

### A. The First Structure: The Graphene-coated Elliptical Nano-wire

In Fig. 2, the configuration of the waveguide has been illustrated. In this structure, a graphene layer has been located on SiO$_2$ elliptical layer, with the permittivity of 2.09 ($\varepsilon_{SiO_2} = 2.09$).

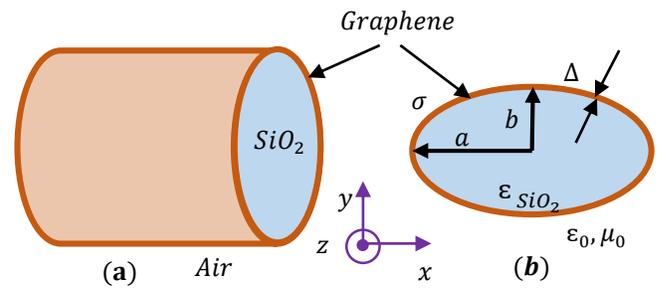

Fig. 2. The graphene-coated elliptical nano-wire: (a) The 3D schematic, (b) The cross-section view.

The long axis and short axis of the ellipse in the simulations are $a = 98\,nm, b = 75\,nm$, respectively. The surrounding medium is assumed to be air. The chemical potential of the graphene is 0.45 eV ($\mu_c = 0.45\ eV$). These parameters are fixed unless otherwise stated. By using the matrix representation of (56), the dispersion relation of the structure is derived.

The effective index ($n_{eff} = Re(k_z/k_0)$) and the propagation length ($L_{prop} = \lambda/4\pi Im[n_{eff}]$) of this waveguide have been depicted in Fig. 3 for 1st and 2nd modes. An excellent agreement is observed between the simulation and theoretical results, which shows the high accuracy of our proposed model. One can see that the effective index increases monotonically, as the frequency increases. While the propagation length decreases gradually with frequency increasing. Another important point is the same behavior of the first two modes for high-frequency range ($f > 40\ THz$). Moreover, it is obvious from Fig. 3 that the 2nd mode has a better performance than the 1st mode for $f < 40\ THz$. As the frequency increases, the plasmonic waves concentrate on the graphene layer, thus the effective index increases. For high frequencies, the propagation loss increases, and then the propagation length reduces.

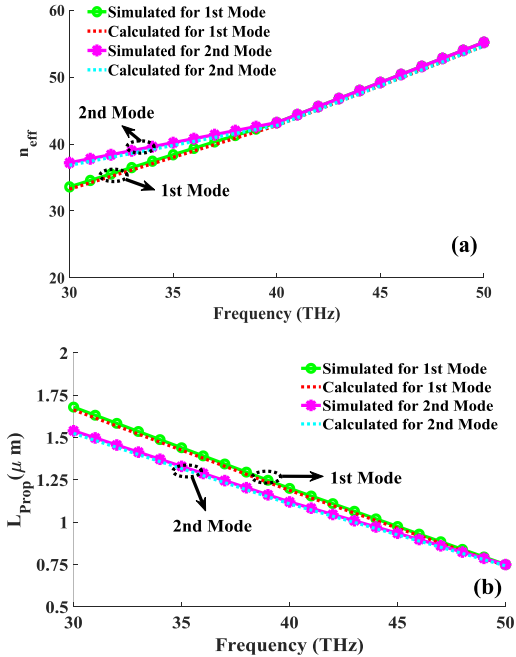

Fig. 3. (a) The effective index, (b) The propagation length of the elliptical nano-wire. The dashed lines show analytical results. The results have been depicted for 1st and 2nd modes.

Fig. 4 illustrates the electric field intensity $|E|$ of the first and second modes for the first structure at the frequency of 40 THz. One can observe from this figure that the electric field has an almost uniform distribution on the graphene-SiO$_2$ interface for the first mode while it has confined at the short axes of the graphene surface for the second mode.

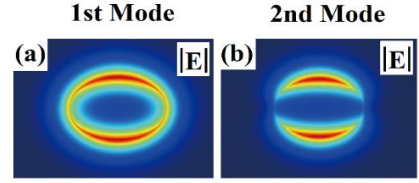

Fig. 4. The electric field intensity $|E|$ of the first structure at the frequency of 40 THz for: (a) The first mode, (b) The second mode.

To study the performance of the plasmonic features, FOM must be defined and investigated. Here, we consider FOM as $Re[n_{eff}]/Im[n_{eff}]$ [52]. Fig. 5 represents FOM as a function of the geometrical parameters of the ellipse for the first two modes. The operation frequency is 40 THz. The short axis in Fig. 5 (a) is 75 nm ($b = 75\ nm$) and the long axis in Fig. 5(b) is supposed to be 98 nm ($a = 98\ nm$). As the long axis ($a$) increases, FOM increases very slowly for 1st mode, while it increases faster for 2nd mode. Moreover, the values of FOM for two modes become equal at a specific long axis ($a = 100\ nm$). This figure clearly indicates that 2nd mode has higher FOM than 1st mode for $a > 100\ nm$. As seen in Fig. 4, the energy has almost regular distribution on the graphene surface for the first mode, whereas it is concentrated at the edges of short axes of the elliptical surface for the second mode. Therefore, as the long axis (a) of the structure increases, the effective index and thus FOM increase slightly for the first mode (see Fig. 5(a)) because the normalized electric field intensity on the graphene surface changes somewhat. However, for the second mode, the field intensity concentrates on the edge of short axes of the elliptical surface by increasing the long axis of the waveguide. As a result, the mode confinement and FOM increase faster for the second mode compared to the first mode, as observed in Fig. 5 (a).

It can be seen from Fig. 5(b) that FOM increases for 1st mode, as the short axis ($b$) increases. While FOM reduces for 2nd mode. For $b < 95\ nm$, the 1st mode has lower FOM than the 2nd mode. As a result, a high value of FOM is obtainable by altering the length of the long and short axis, e.g. FOM= 75 for 2nd mode is achieved for $a = 98\ nm, b = 65\ nm$.

As a final point for this sub-section, the FOM variations as a function of the chemical potential has been demonstrated in Fig. 6. The operation frequency in this diagram is supposed to be 40 THz. It is obvious that high values of FOM can be achieved for the high value of the chemical potential. For instance, FOM of 2nd mode reaches to 90 for $\mu_c = 0.8\ eV$. Furthermore, 2nd mode has better FOM than the 1st mode. Figures 5 and 6 indicate that a superior performance (a high value of FOM) is possible for this waveguide by altering the geometrical parameters $(a, b)$ and the chemical potential $(\mu_c)$.

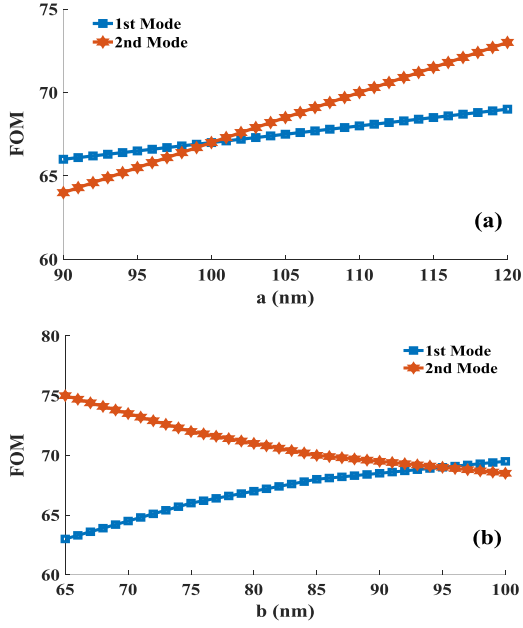

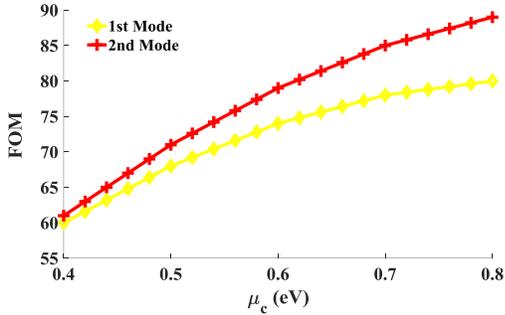

Fig. 5. The analytical results of FOM as a function of: (a) The long axis (a), (b) The short axis (b). The chemical potential is assumed to be 0.45 eV. The frequency is 40 THz. The short axis in Fig. 4 (a) is 75 nm ($b = 75\ nm$) and the long axis in Fig. 4(b) is 98 nm ($a = 98\ nm$).

Fig. 6. The analytical result of FOM as a function of the chemical potential at the operation frequency of 40 THz. The geometrical parameters are $a = 98\ nm, b = 75\ nm$.

*B. The Second Structure: The Graphene-based Elliptical Waveguide with Gyro-electric Layer*

Fig. 7 illustrates a novel graphene-based structure, where a graphene layer has been placed on InSb-SiO$_2$-Si layers. The magnetic bias is applied in the z-direction. The permittivity of Si and SiO$_2$ layers are 11.9 and 2.09 ($\varepsilon_{Si} = 11.9$, $\varepsilon_{SiO_2} = 2.09$), respectively. Here, the gyroelectric layer is supposed to be n-type InSb and its parameters are $\mu_r = \mu_0, \varepsilon_\infty = 15.68$, $m^* = 0.022 m_e, n_s = 1.07 \times 10^{17}/cm^3$, $\nu = 0.314 \times 10^{13} s^{-1}$ and $m_e$ is the electron's mass. Without loss of generality, the surrounding medium is assumed to be air. The chemical potential of the graphene is 0.45 eV. The geometrical parameters are $a_1 = 75\ nm, b_1 = 55\ nm, a_2 = 95\ nm, a_3 = 105\ nm$. Furthermore, $b_2$ and $b_3$ satisfy the following relations:

$$b_2 = \sqrt{a_2^2 - (a_1^2 - b_1^2)} \tag{57}$$

$$b_3 = \sqrt{a_3^2 - (a_1^2 - b_1^2)} \tag{58}$$

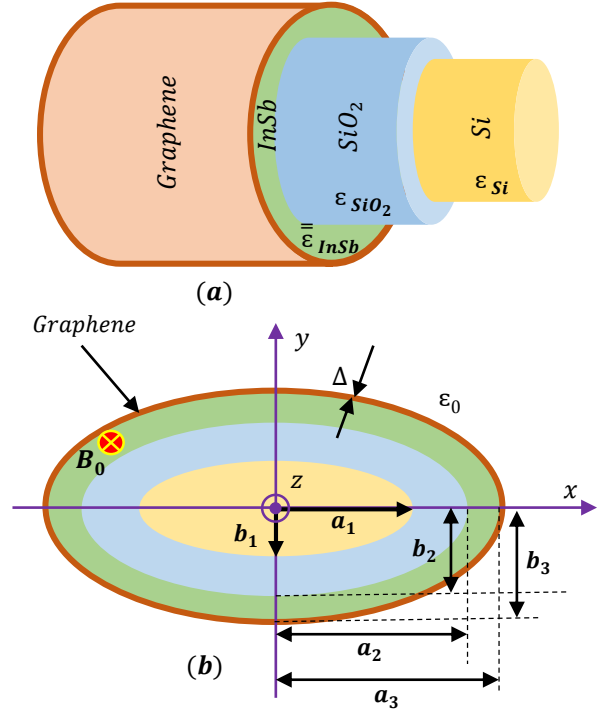

Fig. 7. (a) The 3D schematic, (b) The cross-section of the graphene-based elliptical waveguide with a gyroelectric layer. The external magnetic bias is applied in the z-direction.

By utilizing the analytical model outlined in the previous section, the propagation parameters are derived for this structure. Fig. 8 shows the effective index and the propagation length for various values of external magnetic bias ($B_0 = 0, 1, 2\ T$). For better representation, these parameters are depicted separately for 1$^{st}$ and 2$^{nd}$ modes in different diagrams. There is an excellent agreement between simulation and theoretical results for 1$^{st}$ mode in Fig. 8(a), (b), which validates our proposed analytical model. After this, we will show only analytical results in the remaining diagrams, due to the high accuracy of our proposed model. For this reason, only the analytical results have been depicted for 2$^{nd}$ mode in Fig. 8 (c), (d).

It is evident from Fig. 8 that the effective index increases as the frequency increases, while the propagation length reduces. This matter happens because as the frequency increases, the conductivity of graphene decreases. The magnetic bias has a negligible effect on the modal properties of 2$^{nd}$ mode in this waveguide. As emphasized before, one of the main features of our proposed structure is its ability to control and alter the plasmonic properties by changing the magnetic bias. As seen in Fig. 8, the mode confinement increases drastically for $B_0 = 2\ T$ for 1$^{st}$ mode. Furthermore, this waveguide supports non-reciprocal GPPs, which means that their propagation features change as the direction of the magnetic bias is reversed.

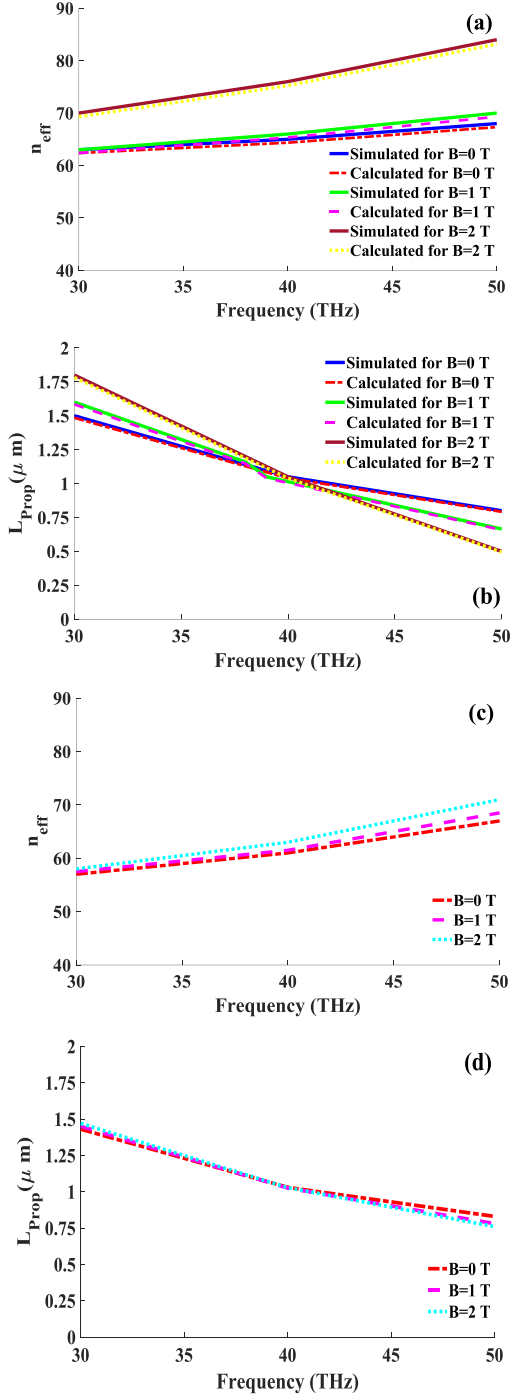

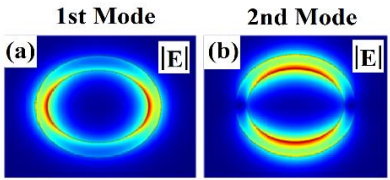

Fig. 9. The electric field intensity $|E|$ of the second structure at the frequency of 40 THz for: (a) The first mode, (b) The second mode.

Fig. 8. The effective index and the propagation length of two modes for various magnetic bias ($B_0 = 0, 1, 2\ T$): (a), (b) 1st mode, (c), (d) 2nd mode. The solid lines in (a) and (b) indicate the simulation results. The dashed lines show the analytical results. The chemical potential is 0.45 eV.

To see the electric field intensity of plasmonic modes for this structure, the electric field intensities $|E|$ of the first and second modes have been shown in Fig. 9 at the frequency of 40 THz. It can be seen that the electric field has a non-uniform intensity for the first mode due to the existence of the gyro-electric substrate. The most energy has been confined at the graphene-InSb border.

To investigate the influence of the geometrical parameters on model properties, it is better to perform a parametric study. Here, we study the effect of $a_1, b_1$ variations on the effective index. Fig. 10 represents the analytical results of the effective index as functions of the long-axis ($a_1$) and short-axis ($b_1$) of Si layer at the operation frequency of 40 THz. The external magnetic bias is assumed to be 1 T. As the long axis ($a_1$) increases, the effective index for two modes reduces. Furthermore, 1st mode has higher mode confinement than 2nd mode, as seen in Fig. 10 (a). One can observe from Fig. 10(b) that the effective index increases for both two modes, as the short axis ($b_1$) increases. The 1st mode has better mode confinement than 2nd mode in this diagram.

We investigate the dependence of FOM on the chemical potential of the graphene in Fig. 11 at the frequency of 40 THz for $B_0 = 1\ T$. For the chemical range of $\mu_c < 0.65\ eV$, two modes approximately have similar diagrams. While 1st mode has better performance (higher FOM) for high values of chemical potential, i.e. the range of $\mu_c > 0.65\ eV$. It can be seen from this figure that a high value of FOM, e.g. FOM=110 for $\mu_c = 0.9\ eV$, is achieved here.

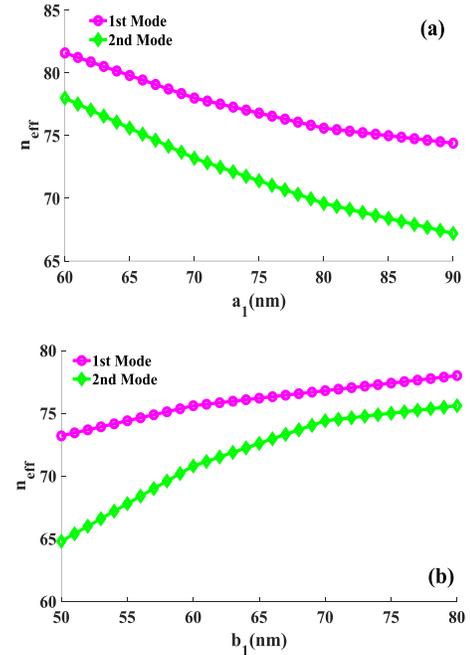

Fig. 10. The analytical results of the effective index as a function of: (a) The long axis (a₁) of Si layer, (b) The short axis (b₁) of Si layer. The frequency is supposed to be 40 THz. In Fig. 8(a), b₁ is supposed to be 55 nm. In Fig. 8(b), a₁ is assumed to be 75 nm. The magnetic bias is 1 T and the chemical potential is 0.45 eV.

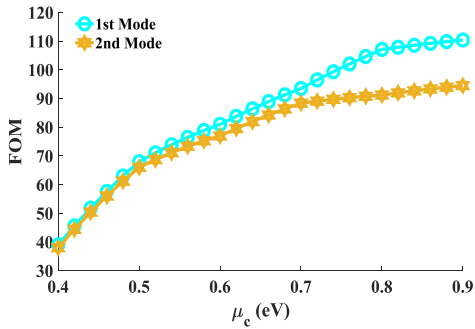

Fig. 11. The analytical results of FOM as a function of the chemical potential at the operation frequency of 40 THz. The external magnetic bias is 1 T. Other parameters are remained fixed.

As a final point, it is worthwhile to be mentioned that the fabrication method of the presented waveguides is similar to the fabrication of graphene-coated nanowires [53-58]. For instance, consider the first structure, i.e. the graphene-coated elliptical nano-wire. To fabricate this structure, first, the $SiO_2$ nanowire should be located on a graphene sheet by using familiar fabrication techniques such as micromanipulation [59]. Second, the structure is pasted to a Silicon-on-Insulator (SOI) wafer and then a nanosecond laser is utilized to cut the graphene sheet along both sides of the elliptical nano-wire [53-55]. Finally, a tapered fiber can be used to eliminate the elliptical nano-wire from the substrate [53-55]. Thus, a graphene-coated elliptical nano-wire is fabricated.

## IV. CONCLUSION

In this paper, a novel theoretical model has been proposed for anisotropic multi-layer elliptical waveguides containing graphene layers. To show the richness of the general structure, two derived specific waveguides have been investigated in this regard. The first, familiar, structure is a graphene-coated elliptical nano-wire showing the familiar propagating properties. The second exemplary case of the proposed general structure is a novel waveguide, constituting air-graphene-InSb-$SiO_2$-Si layers, where the gyroelectric material generates hybrid plasmons, with tunable plasmonic features by altering external magnetic bias and the chemical potential. A very large value of FOM, e.g. FOM=110 for $\mu_c = 0.9\,eV, B_0 = 1\,T$, is obtained for the second structure. Our study is based just on the first two propagating modes as already discussed. A very good agreement between the theoretical and simulation results is observed. In elliptical structures, having two geometric design parameters, the long axis and short axis of each ellipse $(a_1, b_1, a_2, b_2, ...)$, more degrees of freedom are available to the designer. Hence, compared to graphene-based cylindrical waveguides, a superior performance (a high value of FOM) is achievable in these structures. The presented model can be utilized to design tunable THz devices such as cloaks.

## APPENDIX

In section II, writing the electromagnetic fields inside the anisotropic regions and applying boundary conditions led to a matrix representation ($\bar{\bar{S}}$). Due to the elliptical configuration of our general structure and also usage of anisotropic materials, obtaining all elements of the final matrix given in (56) is rigorous. Some elements of this matrix are presented below:

$$S(1,10) = T_{4,2}\, Ne_m(\xi, q_4)\, ce_m(\eta, q_3) \tag{59}$$

$$S(3,4) = Jo_m(\eta, q_2) - \sigma R_{2,1,1}^{Je} \tag{60}$$

$$S(4,8) = R_{3,2,2}^{No} - \sigma T_{3,2}\, Ne_m(\xi, q_3) \tag{61}$$

To obtain the above relations, the boundary conditions defined in (52)-(53) have been used. Indeed, these relations have been derived by substituting the electromagnetic components in (52)-(53) and doing some mathematical procedures. Due to the existence of Q-matrices in the transverse components of electromagnetic fields (see relations (46)-(49)), Q-matrices have appeared inside the above relations, where the following relations have been utilized in (59)-(61) and other elements of the complicated matrix $\bar{\bar{S}}$:

$$\begin{pmatrix} R_{i,N,1}^{Je} \\ R_{i,N,2}^{Je} \end{pmatrix} = \frac{1}{h} \begin{bmatrix} \bar{\bar{\mathbf{Q}}}_N^{Pos} \cdot \begin{pmatrix} T_{i,N} Je'_m(\xi, q_i)\, ce_m(\eta, q_i) \\ Jo'_m(\xi, q_i)\, se_m(\eta, q_i) \end{pmatrix} + \\ j\bar{\bar{\mathbf{Q}}}_N^{Neq} \cdot \begin{pmatrix} T_{i,N} Je_m(\xi, q_i)\, ce'_m(\eta, q_i) \\ Jo_m(\xi, q_i)\, se'_m(\eta, q_i) \end{pmatrix} \end{bmatrix} \tag{62}$$

$i = 2N-1, 2N \quad ; \quad N = 1,2,3,...$

$$\begin{pmatrix} R_{i,N,1}^{Jo} \\ R_{i,N,2}^{Jo} \end{pmatrix} = \frac{1}{h} \begin{bmatrix} \bar{\bar{\mathbf{Q}}}_N^{Pos} \cdot \begin{pmatrix} T_{i,N} Jo'_m(\xi, q_i)\, se_m(\eta, q_i) \\ Je'_m(\xi, q_i)\, ce_m(\eta, q_i) \end{pmatrix} + \\ j\bar{\bar{\mathbf{Q}}}_N^{Neq} \cdot \begin{pmatrix} T_{i,N} Jo_m(\xi, q_i)\, se'_m(\eta, q_i) \\ Je_m(\xi, q_i)\, ce'_m(\eta, q_i) \end{pmatrix} \end{bmatrix} \tag{63}$$

$i = 2N-1, 2N \quad ; \quad N = 1,2,3,...$

$$\begin{pmatrix} R_{i,N,1}^{Ne} \\ R_{i,N,2}^{Ne} \end{pmatrix} = \frac{1}{h} \begin{bmatrix} \bar{\bar{\mathbf{Q}}}_N^{Pos} \cdot \begin{pmatrix} T_{i,N} Ne'_m(\xi, q_i)\, ce_m(\eta, q_i) \\ No'_m(\xi, q_i)\, se_m(\eta, q_i) \end{pmatrix} + \\ j\bar{\bar{\mathbf{Q}}}_N^{Neq} \cdot \begin{pmatrix} T_{i,N} Ne_m(\xi, q_i)\, ce'_m(\eta, q_i) \\ No_m(\xi, q_i)\, se'_m(\eta, q_i) \end{pmatrix} \end{bmatrix} \tag{64}$$

$i = 2N-1, 2N \quad ; \quad N \geq 2$

$$\begin{pmatrix} R_{i,N,1}^{No} \\ R_{i,N,2}^{No} \end{pmatrix} = \frac{1}{h} \begin{bmatrix} \bar{\bar{\mathbf{Q}}}_N^{Pos} \cdot \begin{pmatrix} T_{i,N} No'_m(\xi,q_i) se_m(\eta,q_i) \\ Ne'_m(\xi,q_i) ce_m(\eta,q_i) \end{pmatrix} + \\ j\bar{\bar{\mathbf{Q}}}_N^{Neq} \cdot \begin{pmatrix} T_{i,N} No_m(\xi,q_i) se'_m(\eta,q_i) \\ Ne_m(\xi,q_i) ce'_m(\eta,q_i) \end{pmatrix} \end{bmatrix}$$

$i = 2N-1, 2N \quad ; \quad N \geq 2$ (65)

$$\begin{pmatrix} W_{i,N,1}^{Je} \\ W_{i,N,2}^{Je} \end{pmatrix} = \frac{1}{h} \begin{bmatrix} -j\bar{\bar{\mathbf{Q}}}_N^{Neq} \cdot \begin{pmatrix} T_{i,N} Je'_m(\xi,q_i) ce_m(\eta,q_i) \\ Jo'_m(\xi,q_i) se_m(\eta,q_i) \end{pmatrix} + \\ \bar{\bar{\mathbf{Q}}}_N^{Pos} \cdot \begin{pmatrix} T_{i,N} Je_m(\xi,q_i) ce'_m(\eta,q_i) \\ Jo_m(\xi,q_i) se'_m(\eta,q_i) \end{pmatrix} \end{bmatrix}$$

$i = 2N-1, 2N \quad ; \quad N = 1,2,3,...$ (66)

$$\begin{pmatrix} W_{i,N,1}^{Jo} \\ W_{i,N,2}^{Jo} \end{pmatrix} = \frac{1}{h} \begin{bmatrix} -j\bar{\bar{\mathbf{Q}}}_N^{Neq} \cdot \begin{pmatrix} T_{i,N} Jo'_m(\xi,q_i) se_m(\eta,q_i) \\ Je'_m(\xi,q_i) ce_m(\eta,q_i) \end{pmatrix} + \\ \bar{\bar{\mathbf{Q}}}_N^{Pos} \cdot \begin{pmatrix} T_{i,N} Jo_m(\xi,q_i) se'_m(\eta,q_i) \\ Je_m(\xi,q_i) ce'_m(\eta,q_i) \end{pmatrix} \end{bmatrix}$$

$i = 2N-1, 2N \quad ; \quad N = 1,2,3,...$ (67)

$$\begin{pmatrix} W_{i,N,1}^{Ne} \\ W_{i,N,2}^{Ne} \end{pmatrix} = \frac{1}{h} \begin{bmatrix} -j\bar{\bar{\mathbf{Q}}}_N^{Neq} \cdot \begin{pmatrix} T_{i,N} Ne'_m(\xi,q_i) ce_m(\eta,q_i) \\ No'_m(\xi,q_i) se_m(\eta,q_i) \end{pmatrix} + \\ \bar{\bar{\mathbf{Q}}}_N^{Pos} \cdot \begin{pmatrix} T_{i,N} Ne_m(\xi,q_i) ce'_m(\eta,q_i) \\ No_m(\xi,q_i) se'_m(\eta,q_i) \end{pmatrix} \end{bmatrix}$$

$i = 2N-1, 2N \quad ; \quad N \geq 2$ (68)

$$\begin{pmatrix} W_{i,N,1}^{No} \\ W_{i,N,2}^{No} \end{pmatrix} = \frac{1}{h} \begin{bmatrix} -j\bar{\bar{\mathbf{Q}}}_N^{Neq} \cdot \begin{pmatrix} T_{i,N} No'_m(\xi,q_i) se_m(\eta,q_i) \\ Ne'_m(\xi,q_i) ce_m(\eta,q_i) \end{pmatrix} + \\ \bar{\bar{\mathbf{Q}}}_N^{Pos} \cdot \begin{pmatrix} T_{i,N} No_m(\xi,q_i) se'_m(\eta,q_i) \\ Ne_m(\xi,q_i) ce'_m(\eta,q_i) \end{pmatrix} \end{bmatrix}$$

$i = 2N-1, 2N \quad ; \quad N \geq 2$ (69)